# Effects of Corner Radius and Knudsen Number on Rarefied Gas Flow in a Lid-Driven Cavity

Peiliang Yan[1], Chuang Wen[2,*]

[1] Department of Engineering, Faculty of Environment, Science and Economy, University of Exeter, Exeter, EX4 4QF, UK

[2] Advanced Engineering Centre, School of Architecture, Technology and Engineering, University of Brighton, Brighton BN2 4GJ, UK

*Corresponding authors, Email: c.wen@brighton.ac.uk

**Abstract**

Finite corner curvature is commonly omitted when rarefied cavity flows are represented using idealised geometries. This study examines how downstream corner rounding affects lid-driven cavity flow over $0.5 \leqslant Kn \leqslant 5$. Direct simulation Monte Carlo calculations were performed for normalised corner radii of $Rc$ = 0, 0.10 and 0.20. The model was assessed through published benchmark data and a local grid-sensitivity test. Velocity fields, centreline profiles, area-averaged speed, translational temperature, number density and temperature anisotropy were analysed. A single primary vortex persisted across all cases. Corner rounding modified the local turning of the downstream return flow but caused little change in the wider circulation. The root-mean-square differences between the rounded and sharp-cornered centreline velocity profiles remained below 0.007. The area-averaged normalised speed increased in all rounded configurations, with a maximum change of approximately 2.5%. The thermodynamic response showed a stronger dependence on rarefaction. Corner rounding reduced the mean temperature non-uniformity at $Kn$ = 0.5 and 1, but increased it at $Kn$ = 2 and 5. Changes in the mean density deviation remained within approximately 1.3%, while the mean translational-temperature anisotropy varied from about -3% to 2.4%. The results show that the sharp-corner approximation captures the global velocity structure under the conditions studied, but can omit local flow changes and small thermodynamic responses associated with finite corner curvature.

**Nomenclature**

| *Roman symbols* | | $T_{tr}$ | Mean translational temperature [K] |
|---|---|---|---|
| $A_f$ | fluid area of the cavity [m$^2$] | $T_w$ | Wall temperature [K] |
| $A_T$ | Directional translational-temperature anisotropy | $u$ | Mean gas velocity in the x direction [m/s] |
| $c$ | Molecular velocity vector [m/s] | $u_i$ | Mean gas velocity component in direction i [m/s] |
| $c_i$ | Molecular velocity component in direction i [m/s] | $U_{lid}$ | Lid velocity [m/s] |
| $d_{ref}$ | Reference molecular diameter [m] | $\bar{U}^*$ | Area-averaged normalised in-plane speed |
| $E_u$ | Root-mean-square difference in the horizontal velocity along the vertical centreline | $v$ | Mean gas velocity in the y direction [m/s] |
| $E_v$ | Root-mean-square difference in the vertical velocity along the horizontal centreline | $V$ | Fluid volume [m$^3$] |
| $f$ | Molecular velocity distribution function [s$^3$/m$^6$] | $x$ | Spatial coordinates [m] |
| $F_N$ | Number of physical atoms represented by one simulation particle | $y$ | Spatial coordinates [m] |
| $h$ | Numerical thickness of the computational domain [m] | *Greek symbols* | |
| $k_B$ | Boltzmann constant [J/K] | $\lambda_0$ | Molecular mean free path at the initial equilibrium condition [m] |
| $K_n$ | Knudsen number | $\theta$ | Normalised departure of translational temperature from wall temperature |
| $L$ | Cavity side length [m] | $\omega$ | Variable hard sphere temperature exponent |
| $m$ | Molecular mass [kg] | *Superscripts and subscripts* | |
| $n$ | Local number density [m$^{-3}$] | * | Normalised quantity |
| $n_0$ | Initial number density [m$^{-3}$] | *0* | Initial equilibrium condition |
| $N_p$ | Initial number of simulation particles | $i$ | Coordinate direction |
| $N_s$ | Number of common centreline sampling positions | *ref* | Reference condition |
| $Q$ | Binary collision operator [f/s] | *tr* | Translational quantity |
| $r$ | Corner radius [m] | *w* | Wall quantity |
| $R_c$ | Normalised corner radius | *x,y,z* | Cartesian directions |
| $t$ | Time [s] | *Abbreviation* | |
| $T_i$ | Directional translational temperature in direction i [K] | DSMC | Direct simulation Monte Carlo |
| $T_0$ | Initial gas temperature [K] | PICLas | Particle-in-Cell and Laser simulation software |
| $T_{ref}$ | Reference temperature of the molecular model [K] | VHS | Variable hard sphere |

## 1. Introduction

Gas transport in microscale enclosures becomes increasingly sensitive to solid boundaries as the characteristic length decreases or the operating pressure falls. Molecules can then travel over a substantial part of the enclosure before undergoing an intermolecular collision. Gas–surface interactions consequently account for a greater proportion of momentum and energy exchange. Direct molecular analysis has shown that the near-wall non-equilibrium region can be affected by rarefaction, flow gradients and surface curvature [1]. Kinetic models have also demonstrated that the details of molecular collisions influence the prediction of strongly non-equilibrium gas flows [2]. Under these conditions, velocity slip, temperature jump and non-equilibrium molecular distributions may develop, limiting the accuracy of conventional continuum descriptions [3]. A geometric feature that occupies only a small fraction of a cavity may therefore remain comparable with the distance travelled by individual molecules and influence transport beyond its immediate surroundings.

The lid-driven cavity provides a controlled system for examining this sensitivity. A flat moving lid introduces tangential momentum, while the stationary walls redirect the gas and establish an internal circulation. The classical square cavity has long been used as a numerical benchmark because its geometry is simple and its primary vortex and centreline velocity profiles provide clear quantities for comparison [4]. Under rarefied conditions, reduced intermolecular collision frequency weakens the transfer of momentum from the moving lid to the cavity interior, increases wall slip and changes the position of the primary vortex [5]. Non-equilibrium calculations have also identified spatial variations in density, temperature, stress and heat flux within the enclosure [6].

Kinetic approaches have extended the analysis beyond the range of conventional continuum models. Regularised moment equations have shown that several features of rarefied cavity flow cannot be recovered accurately using simple velocity-slip and temperature-jump corrections [7]. Direct simulation Monte Carlo calculations have revealed non-uniform thermal fields and entropy generation in microcavities [8]. Time-relaxed Monte Carlo methods have also been developed for low-speed cavity flows, where the mean flow velocity can be much smaller than the molecular thermal velocity [9]. These studies established the rarefied lid-driven cavity as a demanding test of molecular momentum and energy transport.

Recent investigations have considered more complex wall motions and operating conditions. A cavity driven by four moving walls was used to examine the combined effects of rarefaction and wall velocity on the flow and thermal fields [10]. Oscillatory square cavities containing binary gas mixtures showed that different molecular species can develop different transport responses under the same wall motion [11]. A double-sided oscillatory cavity was later studied over broad ranges of rarefaction, compressibility and forcing frequency [12]. Thermally induced oscillatory flow in a rectangular enclosure has also been resolved using a coupled macroscopic and kinetic description [13]. These studies provide an increasingly detailed account of the effects of mechanical and thermal forcing. The stationary corners, however, are generally represented as exact intersections between straight wall segments.

The shape of the enclosure introduces a separate influence. Rarefied flow in triangular cavities has shown that inclined walls modify the vortex structure and non-equilibrium heat transport [14]. Thermal stress, thermal creep and edge-driven motion have also been examined in triangular enclosures subject to imposed temperature differences [15]. These results show that a rarefied cavity cannot always be characterised by its outer dimensions alone.

Curved boundaries further modify molecular transport. Rarefied cylindrical Couette-flow calculations have shown that curvature changes wall slip and momentum transfer between moving surfaces [16]. Kinetic schemes developed for moving and curved boundaries have confirmed that local wall orientation must be represented carefully in non-equilibrium thermal flows [17]. Calculations for profiled surfaces moving relative to each other have demonstrated corresponding changes in temperature distribution and heat transfer [18]. Rarefied flow through converging and diverging channels also depends on the way in which the changing cross-section redirects molecular trajectories [19]. Ratchet-shaped microchannels provide additional evidence that local surface orientation can reorganise rarefied flow and species transport [20]. These studies establish that curvature affects molecular transport even when the imposed motion and thermodynamic conditions remain unchanged.

The influence of curvature has recently been examined in lid-driven enclosures where curved walls form a substantial part of the boundary. Zhu et al. studied cylindrical cavities with P-shaped and C-shaped cross-sections using the direct simulation Monte Carlo method and the discrete unified gas kinetic scheme [21]. The P-shaped cavity contains a flat moving lid and a stationary boundary that is largely circular, while the C-shaped cavity uses a curved moving lid. The study considered constant and oscillatory wall motion, rarefaction from the slip regime to the free-molecular regime, and incomplete tangential momentum accommodation. The results showed that cavity shape affects average gas velocity, vortex position, wall slip and the thermal field. In these configurations, curvature changes the overall confinement experienced by the gas. The effect of a finite radius restricted to one stationary corner of an otherwise unchanged square cavity was outside the scope of that study.

Gas–surface interaction may further modify the response to curved confinement. Couette–Poiseuille-flow calculations have shown that molecular reflection and surface interaction affect velocity and stress in a rectangular rarefied channel [22]. Experimental measurements have confirmed that tangential momentum and thermal accommodation depend on the combination of gas species and solid surface [23]. Numerical calculations around microscale beams have also demonstrated that gas–surface interaction changes wall slip, pressure and the resulting force [24]. The present investigation holds the wall-reflection model constant so that the effects of local geometry and rarefaction can be examined without introducing surface accommodation as a third control variable.

Local corner rounding has been studied in rarefied open-cavity flows. Jin et al. used the direct simulation Monte Carlo method to examine hypersonic flow over rectangular cavities with rounded separation and reattachment corners [25]. Rounding at the two locations produced different changes in density, pressure and wall heat flux, demonstrating that the location of the geometric modification matters. Related calculations showed that the boundary-layer thickness at the entrance affects the penetration of momentum and energy into an open cavity [26]. These problems are driven by an external high-speed stream, with gas entering and leaving through the cavity mouth. Their responses depend on boundary-layer interaction, separation, reattachment and compressibility. A closed lid-driven cavity has no inlet, outlet or external boundary layer. Its circulation is generated internally by shear from the moving wall. Trends observed in open hypersonic cavities cannot therefore be transferred directly to the present closed microcavity.

A locally rounded square cavity introduces a more restricted geometric disturbance. The square outer dimensions can remain unchanged, the moving lid can retain its complete length, and most of the stationary boundaries can remain straight. Only the junction between the downstream sidewall and the bottom wall is replaced by a circular arc. The geometry also approaches the standard sharp-cornered cavity continuously as the radius decreases. Differences between the resulting flow fields can therefore be related to a local boundary modification without simultaneously changing the overall cavity shape.

At an ideal right angle, the wall-normal direction changes abruptly between the vertical and horizontal boundaries. Along a finite-radius arc, the wall-normal direction changes continuously. Molecules striking different positions on the arc encounter different incidence angles and return to the enclosure along different trajectories. The first response occurs near the modified corner, although rarefied molecular transport may carry its influence farther into the cavity. Recent analysis of highly rarefied corner flow has shown that corner angle, molecular reflection and intermolecular collisions can control the formation and arrangement of recirculating regions [27]. That study concerned an open triangular corner and does not predict the response of the present closed cavity. It nevertheless demonstrates that a local corner condition can influence a macroscopic circulation pattern through ballistic and collisional molecular motion.

The downstream lower corner provides a useful location for examining this mechanism. With the upper lid moving towards the right, gas adjacent to the downstream wall is redirected into the lower return flow. In continuum lid-driven cavities, corner recirculation is closely connected with the local turning of the flow near stationary-wall junctions [4]. Rounding the downstream lower junction changes the path over which the descending gas turns towards the cavity floor. The disturbance may remain confined to the corner, or it may alter the position and strength of the primary vortex.

The spatial extent of this disturbance may also depend on rarefaction. At lower rarefaction, intermolecular collisions distribute locally redirected momentum through neighbouring molecules. As rarefaction increases, longer molecular trajectories provide more direct connections between the rounded surface and distant parts of the enclosure. These transport mechanisms change at different rates. Geometric arguments alone cannot establish whether the response to a fixed radius will increase, decrease or remain approximately unchanged as rarefaction increases.

The associated thermal response requires separate consideration. Rarefied Couette-flow calculations have shown that molecular energy modes and translational energy components may respond differently to shear and molecular relaxation [28]. A scalar temperature can therefore hide part of the local non-equilibrium state. Directional translational temperatures provide a direct measure of thermal anisotropy, while number-density variation records the redistribution of molecules produced by the driven circulation. These quantities complement the primary-vortex position, centreline velocity and area-averaged circulation by connecting the global flow response with the molecular state of the gas.

The geometric question also has a practical modelling consequence. Fabricated microscale boundaries have finite spatial resolution and need not reproduce a mathematical intersection with zero radius. A corner radius may be small compared with the complete cavity length while remaining relevant to molecular transport. Replacing it with an ideal sharp corner simplifies geometry construction and mesh generation, but the resulting modelling error cannot be inferred from the radius-to-length ratio alone. The calculated difference must also exceed the spatial and statistical uncertainty of the molecular simulation before it can be interpreted as a resolved physical effect.

The direct simulation Monte Carlo method is well suited to this task because it represents molecular motion, intermolecular collisions and wall interactions without assuming local thermodynamic equilibrium. Its predictions nevertheless depend on spatial resolution, time-step size, particle population and statistical sampling [29]. These requirements are especially important when the flow differences produced by local corner rounding may be small. The present calculations are performed using the direct simulation Monte Carlo module of PICLas. The PICLas framework combines particle-based methods for rarefied gas and plasma flows and supports particle tracking and collision modelling on unstructured meshes [30]. The software citation follows the reference recommended by the PICLas development project.

Previous research has established the effects of rarefaction in square cavities, extensive curvature in cylindrical cavities and local corner rounding in open hypersonic cavities. The interaction between a finite downstream corner radius and rarefaction remains insufficiently quantified for a closed, shear-driven square microcavity. It remains unclear whether the response to local rounding is restricted to the modified corner or reaches the primary circulation, and whether its magnitude changes as the gas becomes more rarefied. In this study, direct simulation Monte Carlo calculations are used to investigate monatomic argon flow in a lid-driven square microcavity with a locally rounded downstream corner. The square outer dimensions and full moving-lid length are retained, while the corner radius and Knudsen number are varied systematically. Each rounded configuration is compared with the corresponding sharp-corner case using the primary-vortex position, area-averaged circulation, centreline velocity, number-density non-uniformity and directional translational temperatures. The analysis quantifies how a local geometric disturbance is transmitted through a rarefied enclosed gas and determines whether the corner radius must be represented explicitly under the conditions considered.

## 2. Methods

### 2.1. Geometrical configuration

Fig. 1 shows the three cavity geometries considered in this study. Each configuration retained the same square outer dimensions and the same complete flat moving lid. The geometric modification was restricted to the lower corner on the downstream side of the lid motion. At this location, the original intersection between the bottom wall and the sidewall was replaced by a circular arc. The remaining corners and straight wall sections were unchanged.

The cavity side length was 1 μm. The three configurations had normalised corner radii of 0, 0.1 and 0.2, corresponding to a sharp corner and dimensional radii of 100 and 200 nm. The moving-lid length, cavity height and cavity width remained identical in all configurations. The comparison therefore isolated the influence of the local downstream corner without changing the external dimensions of the enclosure.

The coordinate origin was located at the centre of the external square. In the rounded configurations, the circular arc was tangent to the bottom wall and the downstream sidewall. Its centre was located at ($L/2-r$, $-L/2+r$). The arc met the bottom wall at ($L/2-r$, $-L/2$) and the downstream sidewall at ($L/2$, $-L/2+r$). The wall orientation therefore changed continuously along the arc while remaining unchanged elsewhere.

The normalised corner radius was defined as

$$R_c = \frac{r}{L} \tag{1}$$

The fluid area was:

$$A_f = L^2 - \left(1 - \frac{\pi}{4}\right) r^2 \tag{2}$$

The finite radii reduced the fluid area by approximately 0.215% and 0.858% for $Rc = 0.1$ and 0.2, respectively. The external side length remained the common characteristic length, while the actual fluid area of each geometry was used to evaluate domain-averaged quantities.

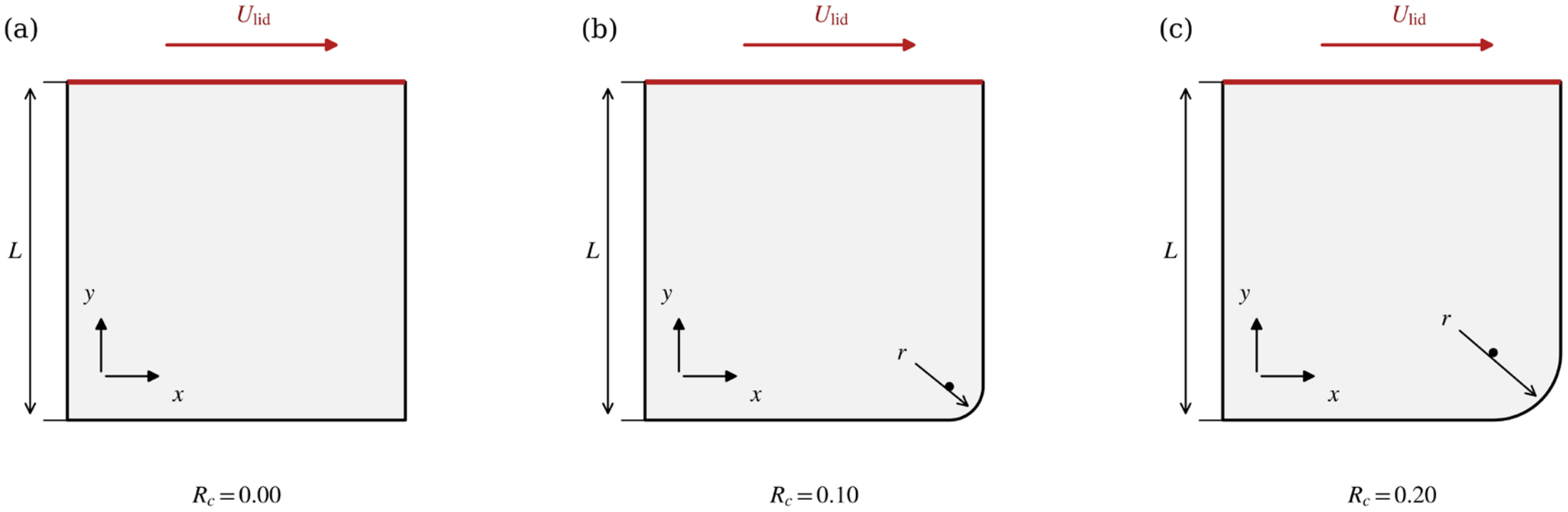

Fig. 1. Geometrical configurations of the lid-driven cavity: (a) $R_c$=0; (b) $R_c$=0.1; and (c) $R_c$=0.2.

## 2.2. Physical model

The rarefied argon flow was calculated using the direct simulation Monte Carlo method implemented in PICLas. The gas was represented by simulation particles, each corresponding to a group of physical argon atoms. Molecular movement and interaction with the cavity boundaries were separated from intermolecular collisions during each time step. Collision partners were selected from particles within the same computational cell, and the macroscopic fields were obtained by accumulating molecular samples over successive time steps.

The molecular velocity distribution is governed by the Boltzmann equation,

$$\frac{\partial f}{\partial t} + c_i \frac{\partial f}{\partial x_i} = Q(f, f) \tag{3}$$

The transport terms describe the temporal and spatial evolution of the molecular distribution, while the collision operator accounts for its redistribution through binary intermolecular collisions. The DSMC method approximates these processes through successive molecular-movement and collision stages.

Argon was treated as a non-reactive monatomic gas. Intermolecular interactions were represented by the variable hard sphere model, and only elastic collisions were included. The molecular model was defined using the reference molecular diameter, reference temperature and temperature exponent listed in Table 1. These parameters remained unchanged throughout the study.

In PICLas, an atomic species is identified by interaction type 1, while collision mode 1 describes elastic collisions without internal-energy relaxation or chemical reactions. The temperature exponent is reported using the convention employed by PICLas and has not been converted to another viscosity-index definition.

Each simulation particle represented a prescribed number of physical argon atoms. The statistical particle weight was calculated from

$$F_N = \frac{n_0 V}{N_p} \tag{4}$$

where $V = A_f h$ for the present extruded cavity. The particle weight was determined separately for each geometry and initial number density. This accounted for the small differences in fluid volume and mesh size while maintaining a similar simulation-particle population per cell.

The rarefaction condition was characterised by

$$Kn = \frac{\lambda_0}{L} \tag{5}$$

The same external cavity side length was used for all geometries. The local corner modification therefore did not change the characteristic length used to define the Knudsen number.

The macroscopic gas properties were obtained from moments of the molecular velocity distribution. The number density and mean velocity were calculated as

$$n = \int f dc,\ u_i = \frac{1}{n}\int c_i f dc \tag{6}$$

The directional translational temperatures were obtained from the second central moments:

$$T_i = \frac{m}{nk_B}\int \left(c_i - u_i\right)^2 f dc,\ T_{tr} = \frac{T_x + T_y + T_z}{3} \tag{7}$$

The corresponding moments were evaluated statistically from the simulation-particle samples accumulated within each cell. The mean translational temperature was used as the gas-temperature measure. The directional temperatures were retained to examine departure from translational equilibrium.

The coordinates, velocity components, translational temperature and number density were normalised as

$$x^* = \frac{x}{L}, y^* = \frac{y}{L}, u^* = \frac{u}{U_{lid}}, v^* = \frac{v}{U_{lid}}, \theta = \frac{T_{tr} - T_w}{T_w}, n^* = \frac{n}{n_0} \tag{8}$$

The temperature departure $\theta$ was used because the moving and stationary walls had the same temperature. Positive and negative values indicate local gas temperatures above and below the common wall temperature.

The horizontal velocity was sampled along the vertical centreline $x^* = 0$, while the vertical velocity was sampled along the horizontal centreline $y^* = 0$. Common normalised sampling positions were used for all cases.

The domain-averaged magnitude of in-plane gas motion was calculated as

$$\overline{U}^{*}=\frac{1}{A_f}\int_{A_f}\frac{\sqrt{u^2+v^2}}{U_{lid}}dA \tag{9}$$

Cell-area weighting was used to evaluate this integral. The numerical thickness was constant, so volume weighting would produce the same value. This quantity measures the mean magnitude of gas motion within the cavity and is not interpreted as circulation.

The departure from directional translational equilibrium was characterised by

$$A_T=\frac{\max\left(T_x,T_y,T_z\right)-\min\left(T_x,T_y,T_z\right)}{T_{tr}} \tag{10}$$

The area-averaged value of $A_T$ was used as the principal measure of translational-temperature anisotropy. Local maximum values were treated as supporting information because isolated extrema are more sensitive to DSMC sampling fluctuations.

The thermodynamic quantities used in the comparison were calculated as

$$\overline{|\theta|}=\frac{1}{A_f}\int_{A_f}|\theta|dA \tag{11.1}$$

$$\overline{|n/n_0-1|}=\frac{1}{A_f}\int_{A_f}|n/n_0-1|dA \tag{11.2}$$

$$\overline{A_T}=\frac{1}{A_f}\int_{A_f}A_T dA \tag{11.3}$$

These quantities measure the mean temperature non-uniformity, number-density non-uniformity and directional translational non-equilibrium, respectively. Each integral was evaluated over the actual fluid area of the corresponding geometry.

The effects of corner rounding on the centreline velocity profiles were quantified using

$$E_u = \left[ \frac{1}{N_s} \sum_{j=1}^{N_s} \left( u_{R_c,j}^* - u_{0,j}^* \right)^2 \right]^{\frac{1}{2}} \tag{12.1}$$

$$E_v = \left[ \frac{1}{N_s} \sum_{j=1}^{N_s} \left( v_{R_c,j}^* - v_{0,j}^* \right)^2 \right]^{\frac{1}{2}} \tag{12.2}$$

Each rounded configuration was compared with the sharp-cornered cavity at the same Knudsen number.

The relative change in the area-averaged normalised in-plane speed was calculated as

$$\delta \overline{U}^* = \frac{\overline{U}^* \left( R_c, Kn \right) - \overline{U}^* \left( 0, Kn \right)}{\overline{U}^* \left( 0, Kn \right)} \times 100\% \tag{13}$$

The relative thermodynamic responses were calculated as

$$\Delta \overline{|\theta|} = \frac{\overline{|\theta|}\left( R_c, Kn \right) - \overline{|\theta|}\left( 0, Kn \right)}{\overline{|\theta|}\left( 0, Kn \right)} \times 100\% \tag{14}$$

$$\Delta \overline{|n/n_0 - 1|} = \frac{\overline{|n/n_0 - 1|}\left( R_c, Kn \right) - \overline{|n/n_0 - 1|}\left( 0, Kn \right)}{\overline{|n/n_0 - 1|}\left( 0, Kn \right)} \times 100\% \tag{15}$$

$$\Delta \overline{A_T} = \frac{\overline{A_T}\left( R_c, Kn \right) - \overline{A_T}\left( 0, Kn \right)}{\overline{A_T}\left( 0, Kn \right)} \times 100\% \tag{16}$$

**Table 1**

Molecular and computational parameters

| Parameter | Symbol | Value |
|---|---|---|
| Cavity side length | $L$ | 1.0 μm |
| Numerical thickness | $h$ | 0.1 μm |
| Lid velocity | $U_{lid}$ | 100 m/s |
| Initial gas temperature | $T_0$ | 300 K |
| Wall temperature | $T_w$ | 300 K |
| Molecular mass | $m$ | $6.6335209\times10^{-26}$ kg |
| Reference molecular diameter | $d_{ref}$ | $4.17\times10^{-10}$ m |
| Reference temperature | $T_{ref}$ | 273 K |
| VHS temperature exponent | $\omega$ | 0.31 |
| Time step | $\Delta t$ | $2.8\times10^{-11}$ s |
| Total number of time steps | - | 40000 |
| Sampling interval | - | 10000 |

**2.3. Computational mesh and sampling procedure**

Fig. 2 shows the meshes used for the three geometries. Each two-dimensional domain was extruded through a numerical thickness of 0.1 μm, with one hexahedral element in the out-of-plane direction. The front and back surfaces were treated as symmetry boundaries. This construction retained three-dimensional molecular velocities while restricting the mean flow to the *x-y* plane.

All meshes consisted of hexahedral cells. The same nominal resolution was used throughout the main part of the cavity. Additional refinement was introduced near the circular arc so that the local curvature could be represented without refining the complete domain. The sharp-cornered, small-radius and large-radius meshes contained 1,024, 1,244 and 1,265 cells, respectively.

The mean initial particle population remained close to 46.5 simulation particles per cell for all three meshes. The different cell counts therefore increased the total number of simulation particles in the rounded geometries without changing the local statistical particle population substantially. No invalid elements, lost particles or mesh-related terminations were detected during the initial tests.

Each formal case was calculated for 40,000 time steps using a fixed time step of $2.8\times10^{-11}$ s. The resulting physical duration was $1.12\times10^{-6}$ s. Macroscopic quantities were accumulated over four consecutive sampling intervals of 10,000 time steps.

The final sampling interval was used for the formal comparisons. Centreline quantities for the three geometries were evaluated at common normalised sampling positions. Domain-averaged quantities were evaluated over the actual fluid region of each geometry.

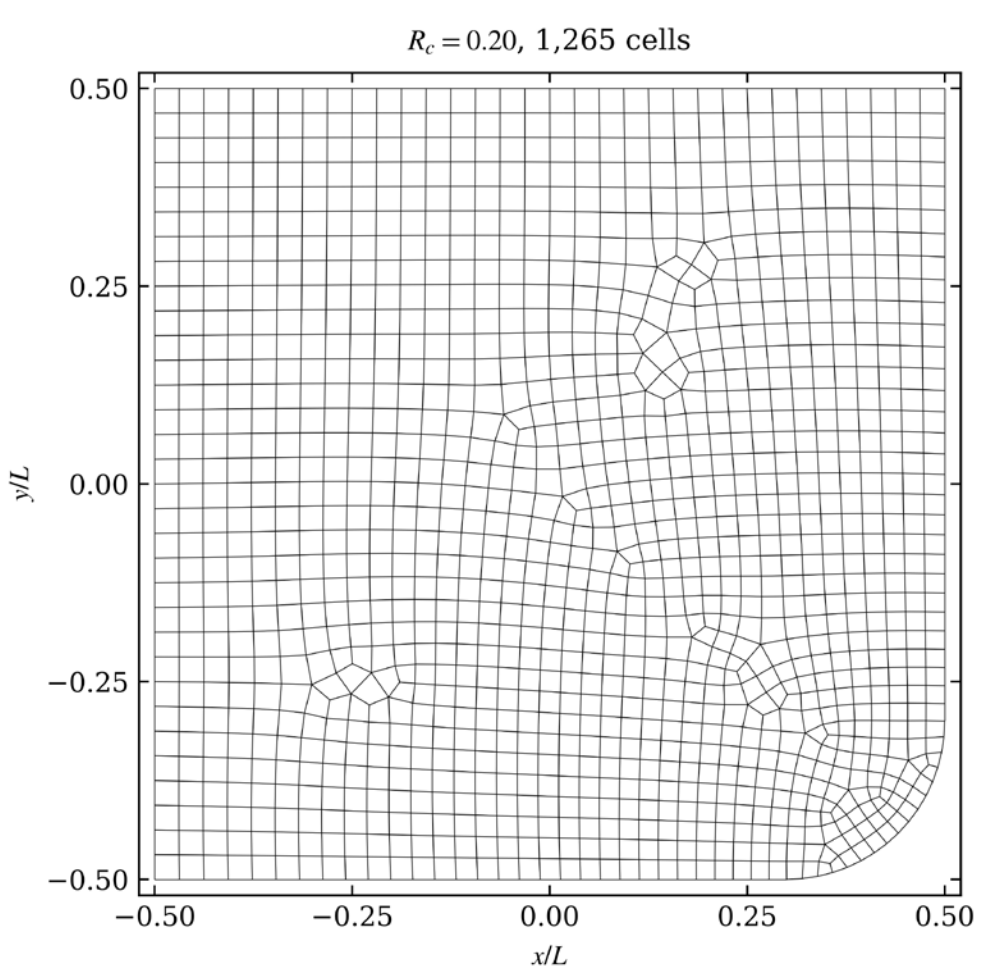

Fig. 2. Computational meshes.

**2.4. Initial and boundary conditions**

The gas was initially stationary and had a uniform temperature and number density. Four initial number densities were selected to produce $Kn$ = 0.5, 1, 2 and 5. The corresponding values were $2.588\times10^{24}$, $1.294\times10^{24}$, $6.470\times10^{23}$ and $2.588\times10^{23}$ $m^{-3}$. The molecular model, initial temperature and cavity dimensions remained unchanged when the number density was varied.

The complete upper wall moved in the positive $x$-direction at 100 m/s. The bottom wall and both sidewalls remained stationary. The gas and all physical walls were initially maintained at 300 K. No temperature difference was imposed between the moving and stationary boundaries.

The solid surfaces were represented by reflective particle boundaries. Complete momentum and translational-energy accommodation were applied throughout the cavity.

In the extended Maxwellian wall model used by PICLas, the momentum accommodation parameter determines the selection between diffuse and specular reflection, while the translational accommodation parameter controls energy exchange between the reflected particles and the wall. Both parameters were set to unity.

The velocity of the moving lid was included when reflected molecular velocities were sampled. The same gas-surface model was applied to every geometry and rarefaction condition. The corner radius and Knudsen number were therefore the only variables in the parameter study.

The three radii and four Knudsen numbers produced 12 formal calculations. Each geometry was evaluated at all four rarefaction conditions, allowing the effect of local corner rounding to be compared at the same Knudsen number.

**2.5. Model validation**

The P-shaped cavity investigated by Zhu et al. [21] was reproduced as an external validation case. The reference study examined rarefied gas flow and heat transfer in cylindrical lid-driven cavities using DSMC and the discrete unified gas kinetic scheme.

The validation was performed at $Kn = 1$ using the normalised vertical velocity along the horizontal centreline. The PICLas result was compared with the same case digitised from the P-cavity DSMC curve. The comparison showed good overall agreement between the PICLas and published DSMC results.

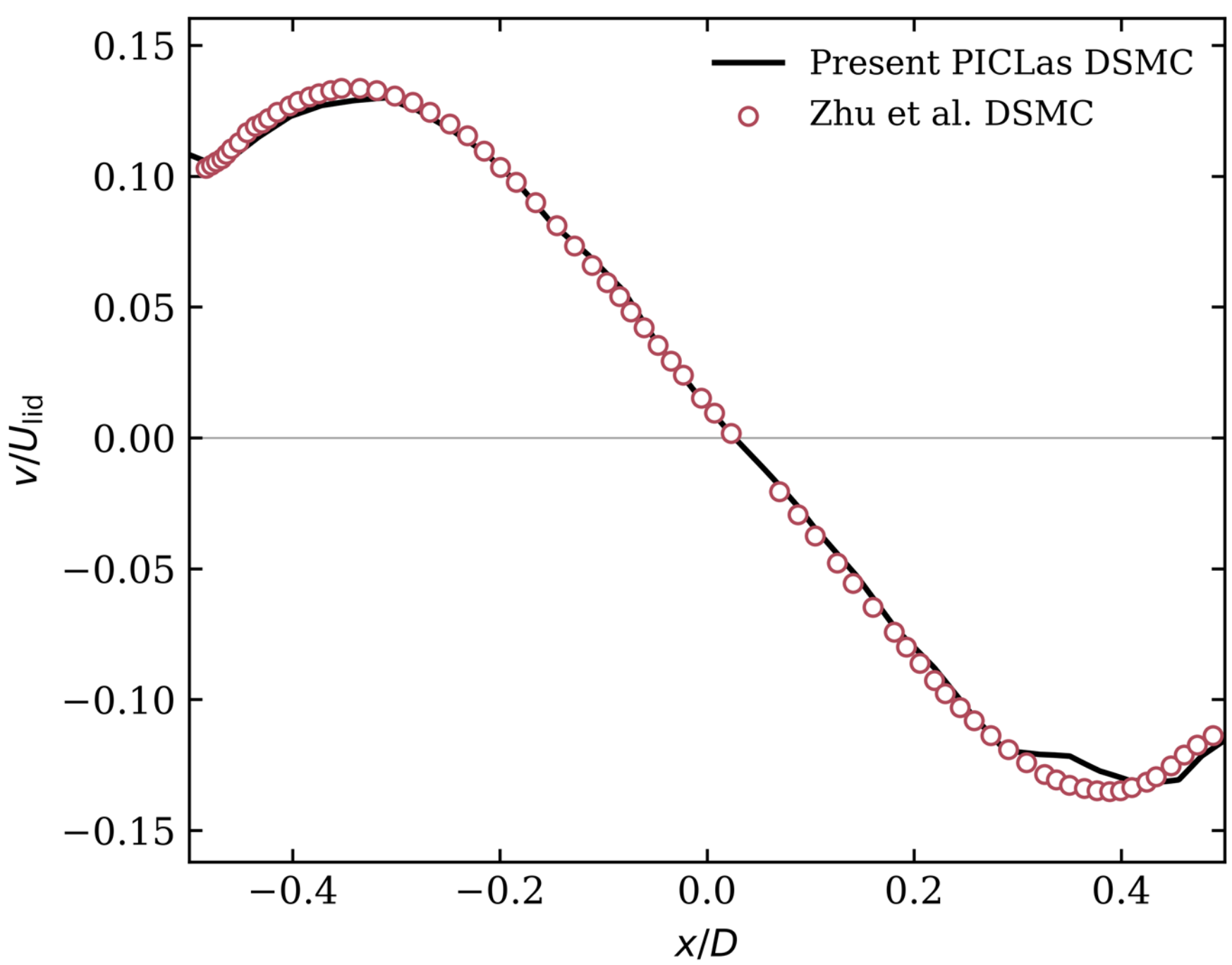

Fig. 3. Comparison of the normalised vertical velocity along the horizontal centreline of the P-shaped cavity at $Kn = 1$ with the published DSMC data.

Grid sensitivity was assessed using the $Rc = 0.1$ geometry because it contained the smallest finite corner radius and imposed the more demanding local curvature-resolution condition. The baseline and refined meshes were compared using the horizontal velocity along the vertical centreline, the vertical velocity along the horizontal centreline and the area-averaged normalised speed. The root-mean-square differences between the two centreline profiles were 0.00427 and 0.00502. The area-averaged normalised speed changed by 0.903%.

The limited changes in both the centreline profiles and the domain-averaged response supported the use of the baseline meshes in the formal calculations.

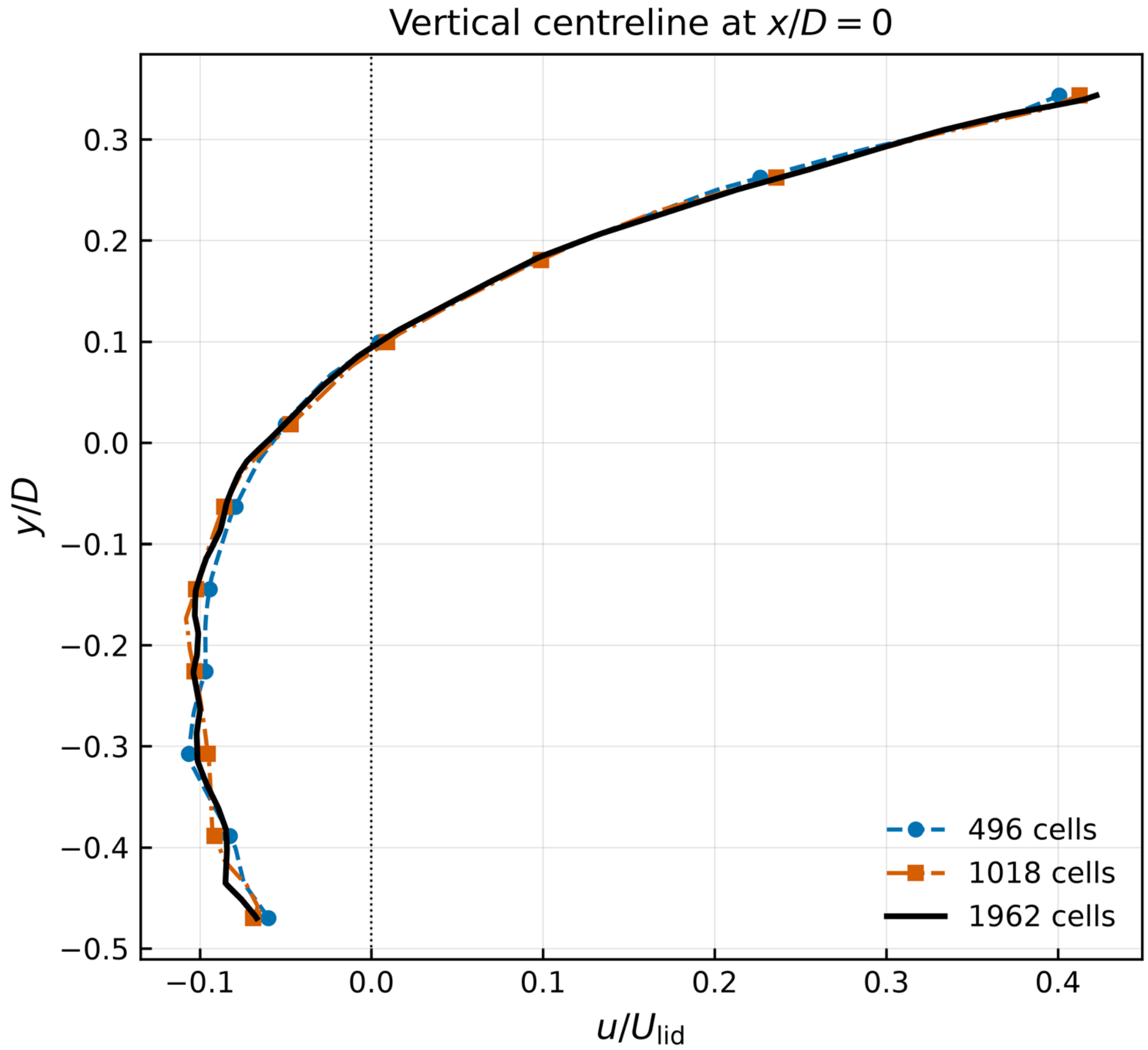

Fig. 4. Grid-sensitivity comparison: vertical velocity along the horizontal centreline.

## 3. Results and discussion

### 3.1. Rarefaction-dependent flow in the sharp-cornered cavity

The sharp-cornered cavity was first examined to establish the rarefaction-dependent flow against which the rounded configurations are compared. As shown in Fig. 5, a single clockwise vortex occupies most of the cavity at all four Knudsen numbers. Gas driven by the moving lid descends along the downstream wall and returns through the lower part of the cavity. The largest velocity magnitude remains confined to a narrow region below the moving lid, while substantially lower velocities occur around the vortex core and near the stationary bottom wall. No distinct secondary corner vortex is resolved over the investigated range of $0.5 \leqslant Kn \leqslant 5$.

The general streamline topology changes only moderately with Knudsen number. The primary vortex remains in the upper half of the cavity, slightly downstream of the vertical centreline. The streamline patterns at *Kn* = 2 and 5 are particularly similar, indicating that the large-scale organisation of the flow changes less at the higher rarefaction levels considered here. Differences remain visible in the shape of the vortex core and in the return flow near the lower and side boundaries.

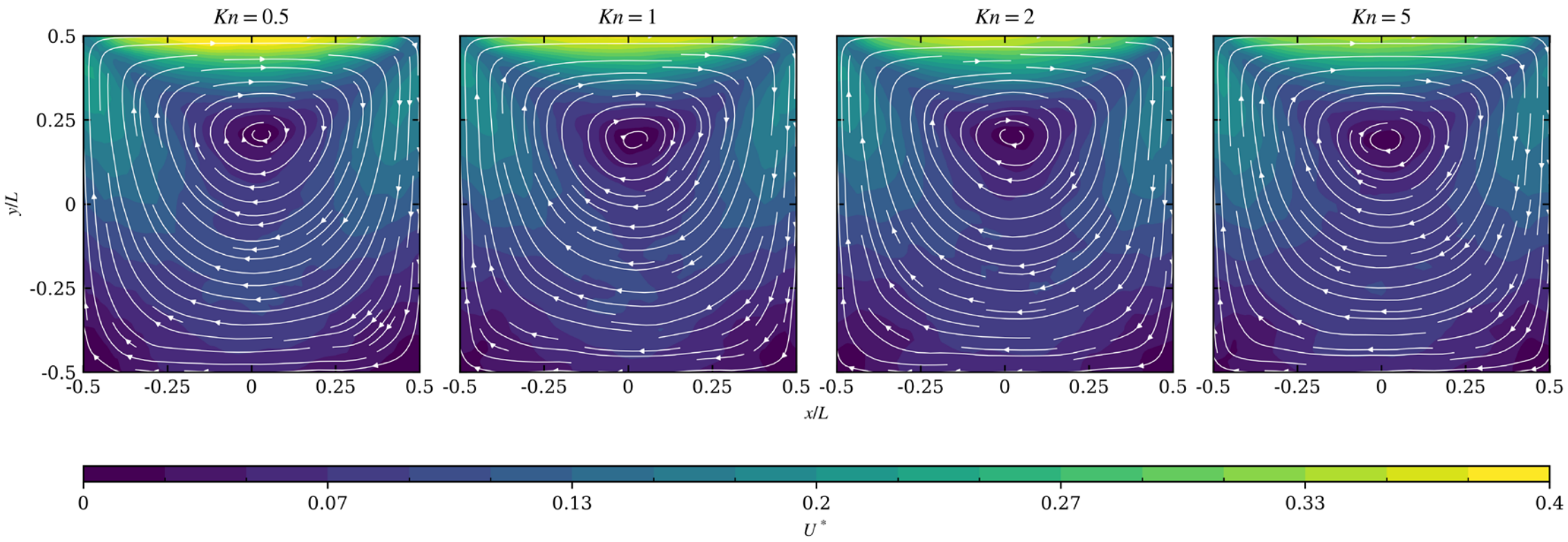

Fig. 5. Normalised velocity magnitude and streamlines in the sharp-cornered cavity at (a) *Kn* = 0.5, (b) *Kn* = 1, (c) *Kn* = 2, and (d) *Kn* = 5.

The centreline profiles in Fig. 6 reveal changes that are less apparent from the streamline patterns. Along the vertical centreline, the horizontal velocity is positive in the upper part of the cavity and negative below the central region. The positive branch represents the penetration of lid-driven momentum, while the negative branch corresponds to the lower return flow. Increasing *Kn* produces a higher gas velocity close to the moving lid but reduces the magnitude of the negative velocity in the lower half. The minimum value changes from approximately -0.10 at *Kn* = 0.5 to about -0.08 at *Kn* = 5. The lower return motion therefore becomes weaker relative to the lid velocity as the gas becomes more rarefied.

Along the horizontal centreline, positive vertical velocity occurs in the upstream half and negative velocity in the downstream half, consistent with the clockwise circulation. The profiles cross zero close to the cavity centre and reach extrema near the two sidewalls. Their magnitudes remain approximately between 0.12 and 0.13 for all four cases. The differences with Knudsen number are modest and not strictly monotonic, particularly near the upstream maximum. Rarefaction therefore changes the distribution of the sidewall-directed motion while preserving the overall balance between the rising and descending branches of the primary vortex.

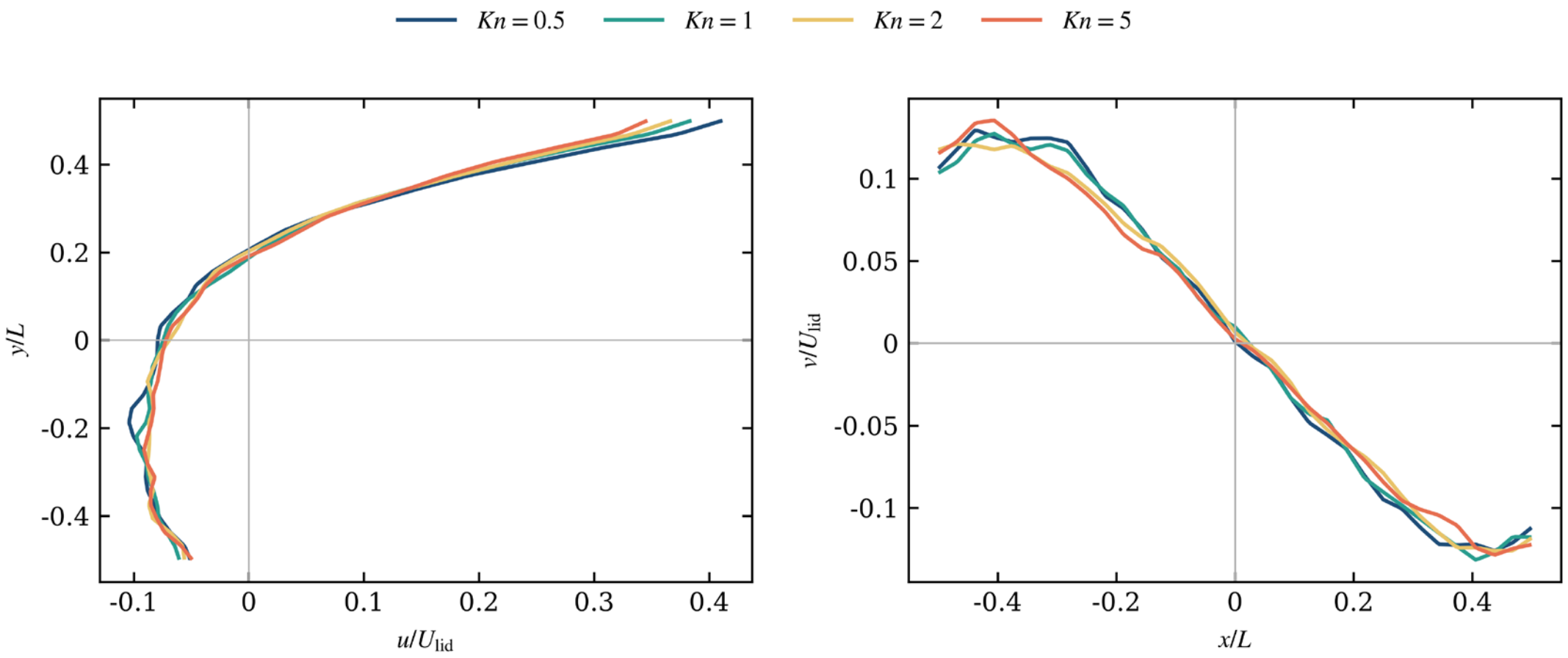

Fig. 6. Normalised centreline velocity along the centreline: (a) horizontal; (b) vertical.

Fig. 7 shows that the mechanically driven flow also produces asymmetric translational-temperature and number-density fields. Although the initial gas and all physical walls are maintained at 300 K, a warmer region develops near the downstream end of the moving lid, while cooler gas occupies much of the upstream side and the lower downstream region. The mean translational-temperature departure remains moderate, with $\theta$ lying mainly between approximately -0.03 and 0.03. Its large-scale spatial arrangement persists across all four Knudsen numbers, although the extent and shape of the warmer and cooler regions vary.

The number density exhibits a clearer upstream-to-downstream redistribution. Values below the initial density occur over the upstream side, while the highest values are concentrated near the downstream wall and upper-right corner. The normalised density extends from approximately 0.82 to 1.15. This variation spans most of the cavity rather than remaining confined to a narrow near-wall region. The overall density pattern remains similar as *Kn* increases, with progressively smoother and more vertically aligned contours through the central region.

Taken together, Figs. 5 to 7 show that rarefaction modifies the strength and internal distribution of the cavity flow without changing its single-vortex topology. The clearest velocity response occurs in the transfer of lid-driven motion towards the lower return flow. The thermal and density fields retain a persistent upstream-to-downstream asymmetry throughout the investigated Knudsen-number range.

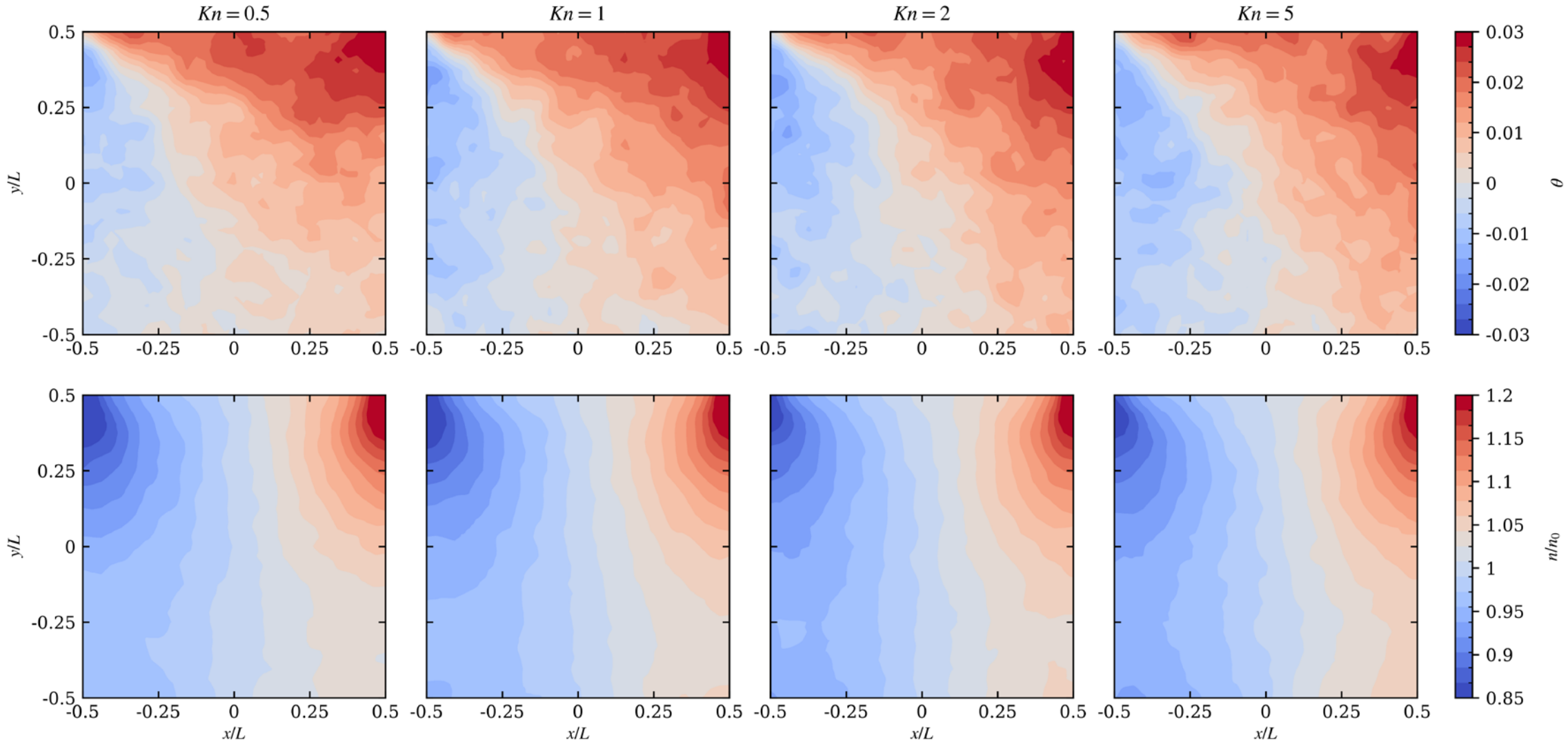

Fig. 7. Translational-temperature departure and normalised number density in the sharp-cornered cavity at *Kn* = 0.5, 1, 2 and 5.

**3.2. Transmission of the corner-induced disturbance**

The influence of corner rounding first appears where the descending flow turns from the downstream wall towards the bottom of the cavity. Fig. 8 compares this region at *Kn* = 1. Replacing the sharp corner with a circular arc changes the available fluid domain and allows the wall direction to vary continuously through the turn. The low-speed region consequently follows the modified boundary, becoming progressively displaced from the original square corner as *Rc* increases. Outside this immediate region, however, the velocity contours and streamline paths remain closely aligned. The finite radius therefore adjusts the local return path without producing a visible reorganisation of the wider circulation.

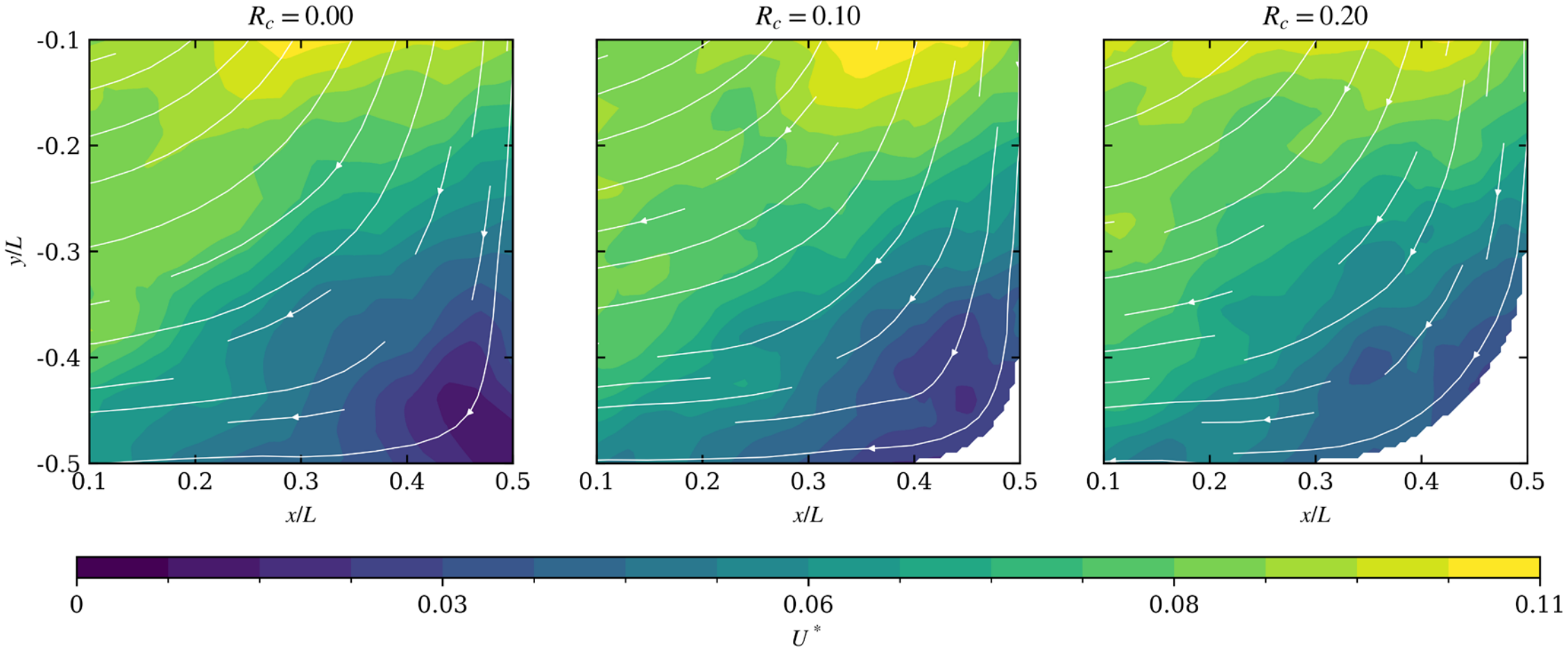

Fig. 8. Normalised velocity magnitude and streamlines in the downstream lower region at *Kn* = 1 for *Rc* = 0, 0.1 and 0.2.

The magnitude of this response is quantified in Fig. 9. The root-mean-square differences between the rounded and sharp-cornered centreline profiles remain below 0.007 for all cases. For the horizontal velocity along the vertical centreline, *Eu* decreases from approximately 0.006 at $Kn = 0.5$ to about 0.004 at $Kn = 1$, before stabilising near 0.005 at the two higher Knudsen numbers. The two rounded geometries produce similar values over most of the investigated range, and the larger radius does not consistently cause a larger centreline difference.

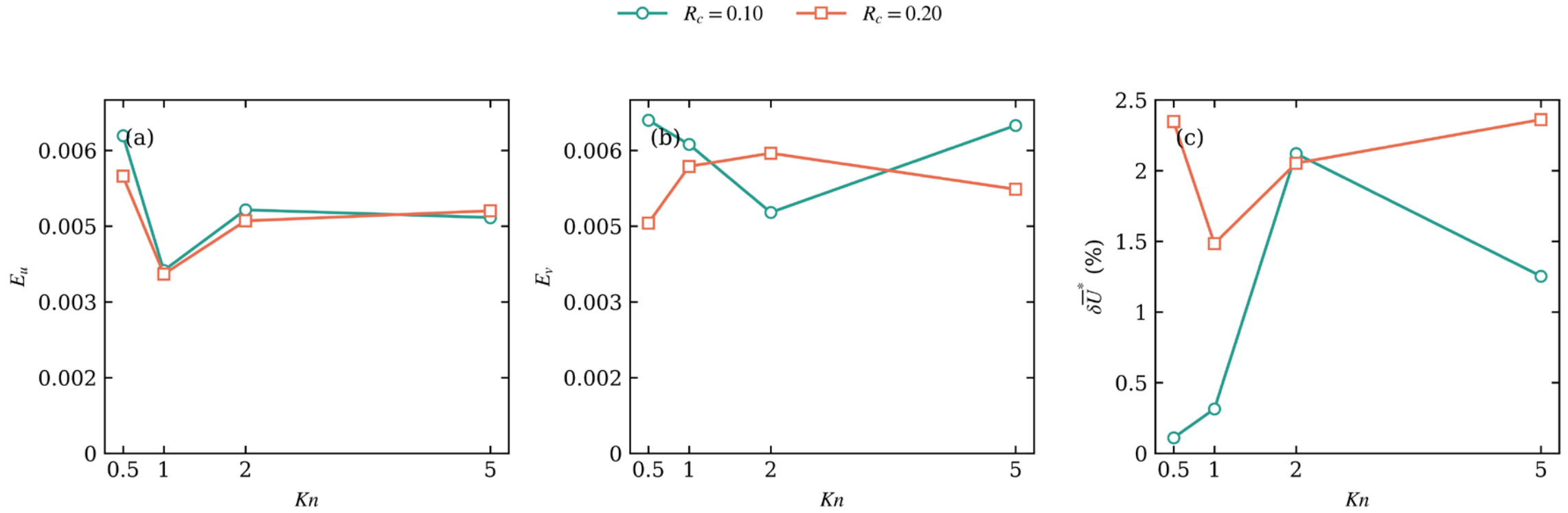

Fig. 9. Velocity responses to downstream corner rounding.

A comparable response is observed for the vertical velocity along the horizontal centreline. The corresponding values of *Ev* remain between approximately 0.005 and 0.0065. Their variation with Knudsen number is non-monotonic and differs between $Rc = 0.1$ and 0.2. These small differences show that the centreline response cannot be described by a simple proportional relationship between corner radius and velocity redistribution. The local geometric change has only a limited influence on the velocity field away from the modified corner.

A clearer dependence on corner radius appears in the area-averaged normalised speed. Corner rounding increases $\bar{U}^*$ in all eight rounded cases. For $Rc = 0.1$, the increase grows from approximately 0.1% at $Kn = 0.5$ to about 2.1% at $Kn = 2$, followed by a reduction to approximately 1.2% at $Kn = 5$. The $Rc = 0.2$ configuration produces a larger increase throughout the investigated range, from approximately 1.5% at $Kn$ = 1 to more than 2% at $Kn$ = 0.5, 2 and 5. The larger radius therefore has a more consistent effect on the domain-averaged motion than on either centreline velocity profile.

Fig. 10 confirms that the global velocity structure remains largely unchanged. For every Knudsen number, the three horizontal-velocity profiles retain similar positive and negative branches, closely matched extrema and nearly unchanged zero-crossing positions. The vertical-velocity profiles also remain closely grouped across the cavity. Small separations occur near the velocity extrema and close to the stationary boundaries, but no systematic displacement of the complete profile is observed. This agreement explains why the streamline topology outside the downstream corner remains similar in Fig. 8.

The local and global measures therefore describe two distinct aspects of the geometric response. Corner rounding modifies the turning of the downstream flow and removes part of the low-speed region associated with the sharp corner. This change produces a small increase in the area-averaged speed, particularly for $Rc$ = 0.2, while leaving the centreline velocity structure largely intact. The primary circulation remains controlled by momentum introduced through the full-length moving lid. Within the present range of $0.5 \leqslant Kn \leqslant 5$, the downstream corner radius mainly influences the local return flow and the overall magnitude of gas motion, with only weak transmission to the cavity centre.

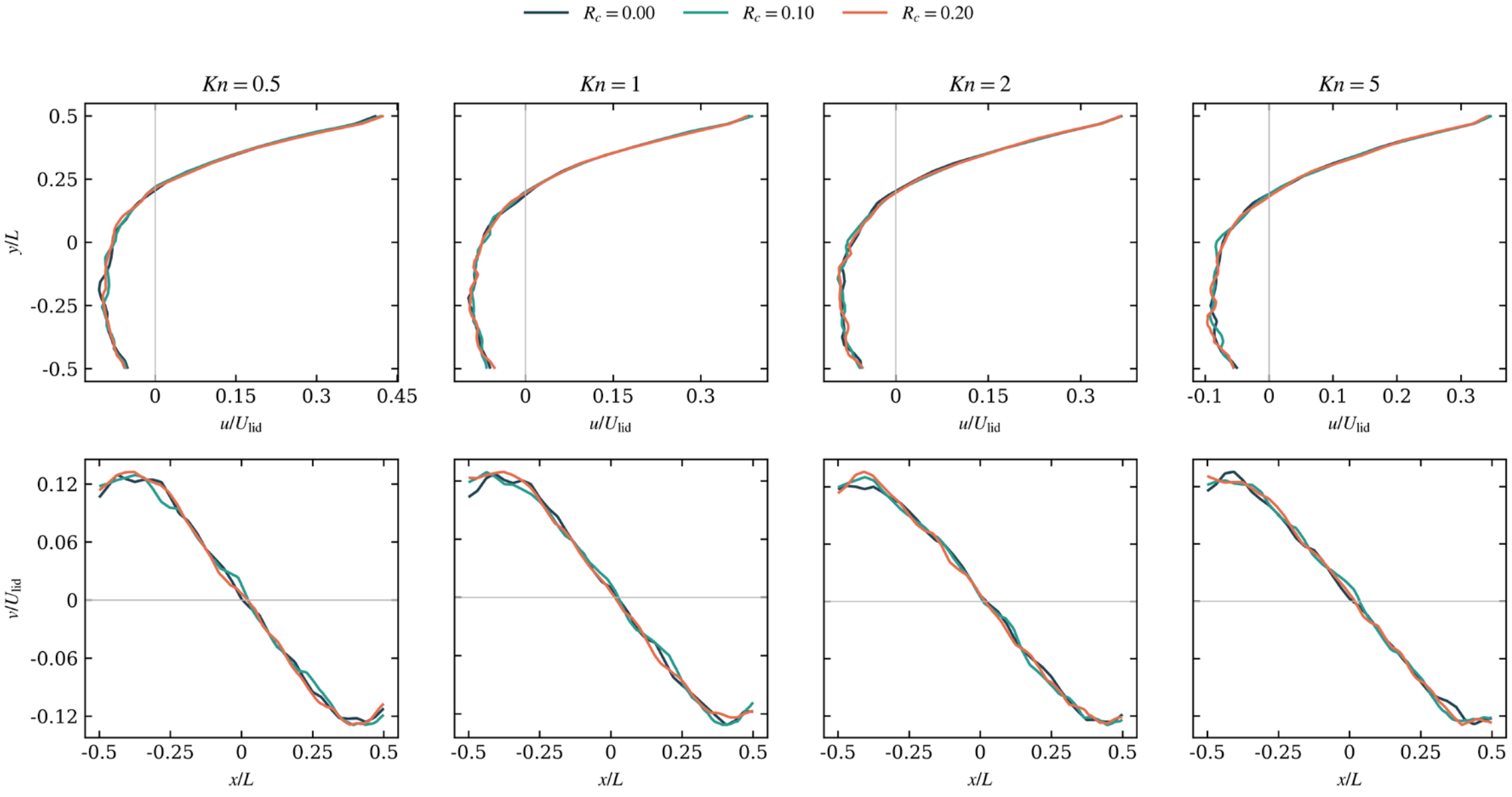

Fig. 10. Centreline velocities at different corner radii and Knudsen numbers.

### 3.3. Thermodynamic response

Fig. 11 examines whether the local geometric modification affects the thermodynamic state of the gas. All quantities are reported as percentage changes relative to the sharp-cornered cavity at the same Knudsen number. This representation separates the effect of corner rounding from the stronger variation caused by rarefaction.

The mean absolute temperature departure in Fig. 11(a) shows a clear change in the direction of the geometric response. At $Kn$ = 0.5, corner rounding reduces $\overline{|\theta|}$ by approximately 7% for $Rc$ = 0.1 and 3% for $Rc$ = 0.2. The reduction remains close to 6% for both radii at $Kn$ = 1. At higher Knudsen numbers, the response becomes positive. The increase is approximately 1% to 2% at $Kn$ = 2, rising to about 3% for $Rc$ = 0.1 and 4% for $Rc$ = 0.2 at $Kn$ = 5. Corner rounding therefore suppresses the overall temperature non-uniformity at the lower Knudsen numbers but enhances it under the more rarefied conditions. The larger radius produces the stronger response at $Kn$ = 1, 2 and 5, although the difference between the two rounded configurations is small at $Kn = 1$ and 2. At $Kn = 0.5$, however, the smaller radius produces the larger reduction.

The density response in Fig. 11(b) is substantially weaker. The change in the mean absolute density deviation remains within approximately ±1.3% for every case. For $Rc = 0.1$, the response changes from a reduction of about 1.3% at $Kn = 0.5$ to an increase of a similar magnitude at $Kn = 1$. The value then decreases to approximately 0.4% at $Kn = 2$ and becomes negative again at $Kn = 5$. The $Rc = 0.2$ configuration produces small positive changes from $Kn = 0.5$ to 2, followed by a reduction of less than 1% at $Kn = 5$. These non-monotonic variations show that the density redistribution is only weakly sensitive to corner radius. Increasing $Rc$ does not consistently increase the density response.

The mean translational-temperature anisotropy in Fig. 11(c) displays a stronger dependence on both radius and Knudsen number. For $Rc = 0.1$, the change remains between approximately −1.1% and 1.2%. The larger radius produces increases of about 2.4% at $Kn = 1$ and 2.0% at $Kn = 2$, followed by a reduction of nearly 3% at $Kn = 5$. The response is therefore non-monotonic, and the sign of the geometric effect depends on the rarefaction condition. In particular, the increase observed at intermediate Knudsen numbers does not persist at $Kn = 5$.

Taken together, the three quantities show that corner rounding produces a measurable but limited thermodynamic response. The strongest systematic change occurs in the temperature non-uniformity, which shifts from a reduction at $Kn \leqslant 1$ to an increase at $Kn \geqslant 2$. Density redistribution is much less sensitive, with changes remaining close to 1%. Temperature anisotropy responds more strongly to the larger radius, although its non-monotonic variation indicates an interaction between local wall geometry and rarefaction.

Combined with the velocity results in Section 3.2, these findings show that downstream corner rounding has only a weak influence on the global flow structure but can modify the thermal and directional non-equilibrium of the gas.

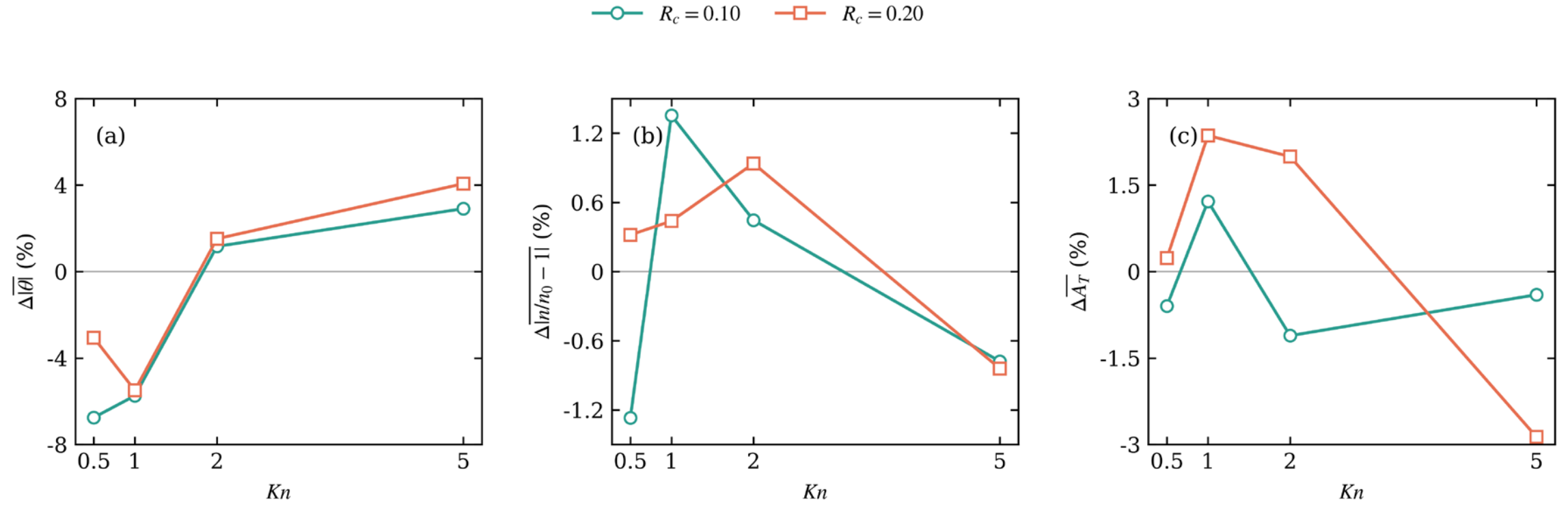

Figure 11. Thermodynamic effects of corner rounding.

## 4. Conclusions

The effects of downstream corner rounding were examined for $Rc = 0$, 0.1 and 0.2 over $0.5 \leqslant Kn \leqslant 5$. Corner rounding altered the local return flow but caused little change in the primary vortex or centreline velocity profiles, whose root-mean-square differences remained below 0.007. The area-averaged normalised speed increased by up to 2.5%, although its variation with corner radius and Knudsen number was not uniformly monotonic.

The thermodynamic response depended more strongly on rarefaction. Corner rounding reduced temperature non-uniformity at $Kn = 0.5$ and 1 and increased it at $Kn = 2$ and 5. Density changes remained within 1.3%, and temperature anisotropy changed by approximately -3% to 2.4%. The sharp-corner approximation therefore captures the global velocity structure under the conditions studied, but may overlook local flow changes and small thermodynamic responses associated with finite corner curvature.